\input lanlmac\overfullrule=0pt
\input labeldefs.tmp
\writedefs
\input mssymb
\def\pre#1{ (preprint {\tt #1})}
%
\input epsf

\long\def\fig#1#2#3{%
\xdef#1{\the\figno}%
\writedef{#1\leftbracket \the\figno}%
\midinsert%
\parindent=0pt\leftskip=1cm\rightskip=1cm\baselineskip=11pt%
\centerline{\epsfbox{#3}}
\vskip 8pt\ninepoint%
{\bf Fig.\ \the\figno:} #2%
\endinsert%
\goodbreak%
\global\advance\figno by1%
}
\font\email=cmtt9
\font\sml=cmss5
\font\ss=cmss10
\def\c{{\hbox{\sml C}}}
\def\t{{\hbox{\sml T}}}
\def\C{{\ss C}}
\def\T{{\ss T}}
\def\D{{\ss D}}
\def\P{{\ss P}}
\long\def\rem#1{}
\def\d{{\rm d}}
\def\der{\partial}

\def\tr{\mathop{\rm tr}\nolimits}
\def\E#1{{\rm e}^{\textstyle #1}}
\Title{\vbox{\baselineskip12pt\hbox{RU-99-13}\hbox{cond-mat/9903385}}}
{\vbox{\centerline{The dilute Potts model}
\vskip2pt\centerline{on random surfaces}}}
\centerline{P.~Zinn-Justin\footnote{$^\dagger$}{{\email
pzinn@physics.rutgers.edu}}}
\medskip\centerline{Department of Physics and Astronomy, Rutgers University}
\medskip\centerline{Piscataway, NJ 08854-8019, USA}
\vskip .3in
\centerline{Abstract}
We present a new solution of
the asymmetric two-matrix model in the large $N$ limit 
which only involves a saddle point analysis. The model
can be interpreted as Ising in the presence of a magnetic field, on random
dynamical lattices with the topology of the sphere (resp.\ the disk) for
closed (resp.\ open) surfaces; we elaborate on the resulting phase diagram.
The method can be equally well applied to a more general $(Q+1)$-matrix
model which represents the dilute Potts model on random dynamical lattices.
We discuss in particular duality of boundary conditions for open
random surfaces.
\Date{03/99}
\nref\BIPZ{E.~Br\'ezin, C.~Itzykson, G.~Parisi and J.-B.~Zuber,
{\it Commun. Math. Phys.} 59 (1978), 35.}
\nref\KAZm{V.A.~Kazakov, {\it Mod. Phys. Lett.} A4 (1989), 2125.}
\nref\IZ{C.~Itzykson and J.-B.~Zuber, {\it J. Math. Phys.} 21 (1980), 411.}
\nref\HC{Harish~Chandra, {\it Amer. J. Math.} 79 (1957), 87.}
\nref\MEHTA{M.L.~Mehta, {\it Comm. Math. Phys.}
79 (1981), 327}
\nref\CHMAME{S.~Chadha, G.~Mahoux and M.L.~Mehta, {\it J. Phys.} A14
(1981), 579.}
\nref\BOULKAZ{D.V.~Boulatov and V.A.~Kazakov, {\it Phys. Lett.} B186
(1987), 379.}
\nref\ALF{J.~Alfaro, {\it Phys. Rev.} D47 (1993), 4714
\pre{hep-th/9207097}.}
\nref\BOUL{D.V.~Boulatov, {\it Mod. Phys. Lett.} A8 (1993), 557
\pre{hep-th/9211064}.}
\nref\STAU{M.~Staudacher, {\it Phys. Lett.} B305 (1993), 332
\pre{hep-th/9301038}.}
\nref\PZJ{P.~Zinn-Justin, {\it Commun. Math. Phys.} 194
(1998), 631\pre{cond-mat/9705044}.}
\nref\PZJK{V.A.~Kazakov and P.~Zinn-Justin, to be published in
{\it Nucl. Phys.} B\pre{hep-th/9808043}.}
\nref\BRG{E.~Br\'ezin and D.J.~Gross, {\it Phys. Lett.} B97 (1980), 120.}
\nref\KOGRN{I.~Kostov, {\it Nucl. Phys.} B (Proc. Suppl.) 10A (1989), 295\semi
D.J.~Gross and M.J.~Newman, {\it Phys. Lett.} B266 (1991), 291.}
\nref\KAZ{V.A.~Kazakov, {\it Nucl. Phys.} B (Proc. Suppl.) 4 (1988), 93.}
\nref\DAUL{J.M.~Daul, preprint {\tt hep-th/9502014}.}
\nref\KSW{V.A.~Kazakov, M.~Staudacher and T.~Wynter,
{\it Commun. Math. Phys.} 179 (1996), 235\pre{hep-th/9506174}.}
\nref\MSS{G.~Moore, N.~Seiberg and M.~Staudacher, {\it Nucl. Phys.}
B362 (1991), 665.}
\nref\KOS{I.~Kostov, {\it Nucl. Phys.} B376 (1992), 
539\pre{hep-th/9112059}.}
\nref\KOSTA{I.~Kostov and M.~Staudacher, {\it Nucl. Phys.} B384
(1992), 459\pre{hep-th/9203030}.}
\nref\FUHASA{K.~Fukazawa, K.J.~Hamada and H.~Sato,
{\it Mod. Phys. Lett.} A5-29 (1990), 2431.}

\newsec{Introduction}
The study of various multi-matrix models in the large $N$ limit
is motivated by their
interpretation as statistical lattice models on random surfaces.
The one-matrix model \BIPZ\ already describes the summation over random
surfaces, but to put ``matter'' on the surface,
several matrices are required\foot{Except
the Kazakov multicritical points \KAZm\ which correspond to non-unitary
matter.}. The simplest such model is
the two-matrix model, which has the following
partition function [\xref\IZ,\xref\MEHTA]
\eqn\twomma{{\rm Z}(\alpha_0,\beta_0,\gamma)=
\int\!\!\!\int \d A\, \d B\, \E{N\tr\left[-{1\over 2}(A^2+B^2)
+{\alpha_0\over 3}A^3+{\beta_0\over3}B^3+ {1\over\gamma}AB\right]}
}
where $A$ and $B$ are $N\times N$ hermitean matrices.
In the large $N$ limit, this generates triangulated surfaces
with the spherical topology, on which spins live (the
two matrices $A$ and $B$ correspond to spins up and down)
\BOULKAZ, thus reproducing the Ising model on random surfaces. Schematically,
$\alpha_0$ and $\beta_0$ play the roles of both magnetic field $H$
and cosmological constant ($\alpha_0/\beta_0=\exp(2H/T)$),
while $\gamma$ is related to the
temperature $T$ by $\gamma=\exp(1/T)$.

This model has been solved using orthogonal polynomials \MEHTA,
but, strangely, the simplest tool available, the saddle point method,
which works so well for the one-matrix model, has not been
used. It is commonly assumed that this method does not work in the
two-matrix model, the usual argument being the following:
after diagonalization of $A$ and
$B$ and use of the Itzykson--Zuber--Harish Chandra formula [\xref\HC,\xref\IZ],
we are left with an integral over the eigenvalues:
\eqn\twommb{
{\rm Z}(\alpha_0,\beta_0,\gamma)=\int\!\!\!\int\prod_i \d
a_i \d b_i \Delta[a_i] \Delta[b_i]
\det_{i,j}\Big[\E{N{1\over\gamma} a_i b_j}\Big]
\E{ N\sum_i\left[-{1\over 2}(a_i^2+b_i^2)
+{\alpha_0\over3}a_i^3+{\beta_0\over3}b_i^3\right]}
}
where $\Delta(\cdot)$ is the Van der Monde determinant.
Then, one uses symmetry of permutation of the eigenvalues to reduce
this expression to:
\eqn\twommc{
{\rm Z}(\alpha_0,\beta_0,\gamma)=\int\!\!\!\int\prod_i \d
a_i \d b_i \Delta[a_i] \Delta[b_i]
\E{ N\sum_i\left[-{1\over 2}(a_i^2+b_i^2)
+{\alpha_0\over3}a_i^3+{\beta_0\over3}b_i^3+{1\over\gamma} a_ib_i\right]}
}
At this stage, a large $N$ saddle point analysis shows that there is a
continuous infinity of saddle points and it is very difficult to
derive anything from it. Of course, the problem stems from the
transformation of \twommb\ to \twommc: since we have broken the symmetry
of permutation of the eigenvalues, each ordering $a_{\sigma(1)}
<\cdots<a_{\sigma(N)}$ of the eigenvalues leads to a different saddle
point, so that we have $N!$ saddle points, which causes trouble as
$N\to\infty$. The problem does not exist at the level of \twommb\ and
in fact, as we shall show, there is a well-defined saddle point to it;
we shall then explain how to determine it, which will
result in a very compact and elegant way of expressing the resolvent(s)
of the model. A similar result could be obtained for the two-matrix
model with quartic vertices, but we shall not choose to do so here.

We shall then generalize our results to the dilute $Q$-states Potts model on
random surfaces, which is defined by 
(see \ref\NBRS{B.~Nienhuis, A.N.~Berker, E.K.~Riedel and
M.~Schick, {\it Phys. Rev. Lett.} 43 (1979), 737.} for a somewhat similar
definition on a flat lattice)
\eqn\Qmma{{\rm Z}_Q(\alpha_0,\beta_0,\gamma)=
\int\!\!\!\int \d A\,\prod_{q=1}^Q \d B_q\,
\E{ N\tr\left[-{1\over 2}(A^2+\sum_{q=1}^Q B_q^2)
+{\alpha_0\over 3}A^3
+\sum_{q=1}^Q ({\beta_0\over3}B_q^3+{1\over\gamma}AB_q)\right]}
}
This model describes the following statistical model on random
surfaces: each vertex of the
surface is either unoccupied (represented by matrix $A$) or occupied
by a spin in $Q$ possible states (matrices $B_q$), with the rule
that adjacent vertices cannot be occupied by spins in different states.
Again, $\alpha_0$ and $\beta_0$ are cosmological constants
and control the dilution (i.e.\ density of
unoccupied sites), and $\gamma$ is related to the inverse temperature. 
The two-matrix model \twomma\ is the particular case $Q=1$. 

From now on we shall redefine the couplings of the model
and rescale the fields so that the partition function can be rewritten
\eqn\Qmm{{\rm Z}_Q(\alpha,\beta,\gamma)=
\int \d A\,
\E{ N\tr\left[-{\gamma\over 2}A^2+{\alpha\over 3}A^3\right]}
\left(\int \d B\, \E{ N\tr\left[-{\gamma\over 2}B^2
+{\beta\over3}B^3+AB\right]}\right)^Q
}
where $\alpha=\alpha_0\gamma^{3/2}$, $\beta=\beta_0\gamma^{3/2}$.
The main physical quantities of the model are the
resolvents, which are defined by
\eqn\defres{\eqalign{
\omega_A(a)&=\left<{1\over N}\tr{1\over a-A}\right>\cr
\omega_B(b)&=\left<{1\over N}\tr{1\over b-B}\right>\cr
}}
where the large $N$ limit is implied and $a$, $b$ are complex numbers.
$\omega_A(a)$ and $\omega_B(b)$ are generating functions of averages
of the form $\left<\tr A^n\right>$ and $\left<\tr B^n\right>$,
but they are also important
from the diagrammatic point of view: they correspond to
sums over connected surfaces with a boundary (``loop functions''),
the parameters
$a$ or $b$ playing the role of boundary cosmological 
constant.
In the large $N$ limit, these surfaces have the topology of a disk.
The difference between $\omega_A(a)$ and $\omega_B(b)$ lies in the
{\it boundary conditions}: for $\omega_A(a)$ (resp.\ $\omega_B(b)$),
there are only matrices $A$ (resp.\ $B$) on the boundary.
The large $n$ asymptotics of $\tr A^n$ and $\tr B^n$
(i.e.\ surfaces with large boundary) are
dominated by the singularities of the corresponding resolvent.
These singularities are also relevant for the physics in the bulk:
if $g$ is the exponent of this singularity, then the central charge $c$ of
the critical model is given by [\xref\MSS,\xref\KOS]
\eqn\cc{c=1-6(\sqrt{g}-1/\sqrt{g})^2}
These resolvents will therefore play a central role in our analysis,
and our goal will be to find exact expressions for them.
Let us remark that when
the dilution is turned off ($\alpha=0$), one can perform
the gaussian integration over $A$ and we are brought back to
the usual $Q$-states Potts model on random surfaces
\KAZ. $\omega_B(b)$ is then the standard
resolvent: it corresponds to boundary conditions of the Potts model
where all the spins on the boundary are in a given state. 
$\omega_A(a)$ is not a natural resolvent, because the sites on the
boundary are unoccupied, whereas they cannot be so in the bulk.
However, once dilution it turned
on, the two resolvents $\omega_A(a)$ and $\omega_B(b)$
should be treated {\it on equal footing}; they will correspond, as
we shall explain later, to
boundary conditions which are ``dual'' to each other.

The plan of the article therefore goes as follows: first we shall
analyze in section 2 the integral over $B$ which is common
to the models \Qmm\ for arbitrary
$Q$, then discuss the cases $Q=1$, $2$ and $3$, $4$
(though in \Qmm\ the parameter $Q$ can take arbitrary real values,
for simplicity we only consider here integer values)  in sections 3,
4, 5 and finally conclude in section 6.

\newsec{The external field problem}
Let us first consider part of the model only: if the matrix $A$ is
held fixed in \Qmm,
we are left with the problem of one matrix in an
external field [\xref\BRG,\xref\KOGRN]:
\eqn\extf{
\Xi(A)=\int \d B\, \E{ N \tr [-V(B) + AB]}
}
where $V$ is a polynomial potential. We shall show how the saddle
point equations expressed in \PZJ\ allow indeed to calculate
this matrix integral, and how in the case of the cubic potential
$V(b)=-{\beta\over3}b^3+{\gamma\over2}b^2$, they in fact
reproduce the solution \KOGRN\ without the use of any partial differential
equations.

By $U(N)$-invariance $\Xi(A)$ only depends on the eigenvalues $a_i$ of
$A$:
\eqn\extfb{
\Xi[a_i]=\int \prod_i \d b_i \Delta[b_i] 
{{\displaystyle \det_{i,j}}\big[\E{Na_i b_j}\big]\over\Delta[a_i]}
\E{ -N\sum_i V(b_i)}
}
We shall now show how to calculate in the large $N$ limit
the logarithmic derivatives
of $\Xi$ with respect to $a_i$ ($\Xi$ then follows by simple integration).

We assume that the density of eigenvalues of $A$ becomes smooth in the
large $N$ limit, and denote it $\rho_A(a)$; for the sake of simplicity only,
its support will be taken to be of the form of a single interval
$[a_1,a_2]$. It is related to the resolvent of $A$ by
\eqn\resA{\omega_A(a)=\int_{a_1}^{a_2} {\d a' \rho_A(a')\over a-a'}}
which is an analytic function everywhere except on the support of $A$,
where it has a cut (leading, if $\rho_A(a)$ is smooth, 
to other sheets) which we call the physical cut.

We can now write \PZJ
\eqn\derlog{{1\over N}{\der\over\der a_i} \log \Xi[a_i]=b(a_i)-\omega_A(a_i)}
where $b(a)$ is an analytic (multi-valued) function which has the same
physical cut as $\omega_A(a)$. The particular sheet corresponding to the value
of $b(a)$ in \derlog\ is called the physical sheet. We see that
$b(a)$ is the quantity we need to compute.

In order to do so, we must write down the saddle point equation of
\extf, which we shall do carefully. First we introduce symmetrically
the resolvent
$\omega_B(b)$ of $B$, with a physical cut $[b_1,b_2]$ (i.e.\ the
eigenvalues of $B$ fill the interval between $b_1$ and $b_2$):
$$\omega_B(b)=\int_{b_1}^{b_2} {\d b'\rho_B(b')\over b-b'}$$
Taking the logarithmic derivative with respect to the $b_i$
results in the appearance of the function $a(b)$, which has
the same cut as
the resolvent $\omega_B(b)$ of $B$, and is the {\it functional
inverse} of $b(a)$ (see appendix 1 of \PZJ). 
The saddle point equations then read:
$${1\over2}(\omega_B(b+i0)+\omega_B(b-i0))
+{1\over2}(a(b+i0)+a(b-i0))=V'(b)\qquad b\in [b_1,b_2]$$
Using the fact that $a(b)$ and $\omega_B(b)$ have the same cut, this equation
can be analytically continued:
\eqn\speb{\omega_B(b)+a_\star(b)=V'(b)}
where $a_\star(b)$ is the value of $a(b)$ on the other side of the physical cut of $B$.
Before going further, let us give the physical significance of $a(b)$. Formally, the
saddle point equation \speb\ can be written down:
${\d\over\d b} {\delta S\over\delta \rho_B(b)}=0$,
where $S$ is the action, and the derivative $\d/\d b$ takes care of the normalization condition
$\int\d b\, \rho_B(b)=1$. Now $\delta S/\delta\rho_B(b)$ is nothing but the {\it effective potential}
$V_{\rm eff}(b)$
for the eigenvalues $b_i$, which must be the sum of three terms: the potential $V(b)$, the effective
interaction among the $b_i$, and the effective interaction between the $b_i$ and the $a_i$.
Indeed, we find explicitly that
$$ V_{\rm eff}(b)=\left\{\eqalign{
{\rm const}\hskip3cm &b\in[b_1,b_2]\cr
V(b)-\int^b \left[\omega_B(b)+a(b)\right] \d b\qquad &b\not\in[b_1,b_2]\cr
}
\right.$$
(note that it is $a(b)$ which appears and not $a_\star(b)$ because we are outside the cut).
Therefore $a(b)$ can be interpreted as the derivative of the effective potential for the action of
the $a_i$ on the $b_i$\foot{Symmetrically, $b(a)$ will be the derivative
of the effective
potential for the action of the $b_i$ on the $a_i$.}.

Let us show now how to handle the equation \speb: it can be rewritten
\eqn\spebb{\omega_B(b(a))+a=V'(b(a))}
where it is understood that \spebb\ is a relation between multi-valued
functions.
To use this equation, we now assume that $V$ is a polynomial of degree
$d$, and look at the behavior of $b(a)$ as $a\to\infty$.
Under minimal assumptions on the analytic structure of $b(a)$ (see below), on all cuts connected
to the physical sheet without crossing the physical cut, $b(a)\to\infty$ as $a\to\infty$.
Therefore $b(a)^{d-1}\sim a$ as $a\to\infty$, which leads us to the ``minimal'' conjecture
for the analytic structure of $b(a)$ which is depicted by figure
\minconj\ (a similar structure was found for large $N$ characters in \KSW).
All the cuts except the physical cut go to infinity, and we assume that they
do not cross the physical cut\foot{This corresponds to the strong
coupling phase of \BRG. In the present context, it simply means we remain below
the continuum limit surface.}.
\fig\minconj{Analytic structure of $b(a)$ in the external field problem
with a polynomial potential of degree $d$.}{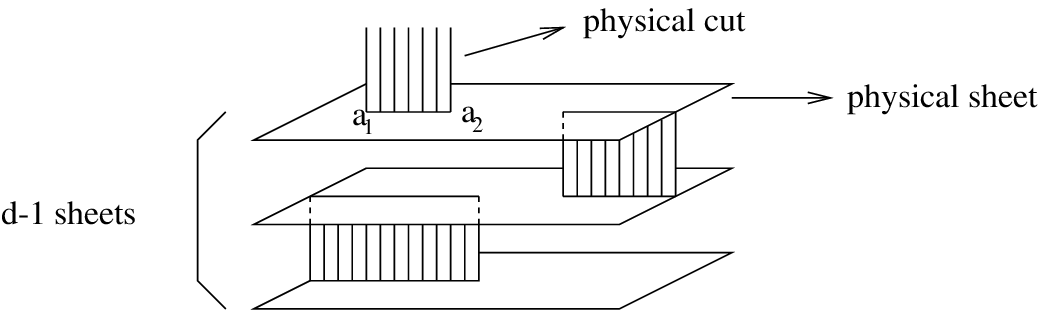}

At this point we have complete knowledge of the analytic behavior of $b(a)$ (the physical cut
is determined by the fact that it is identical to that of $\omega_A(a)$, which is known
by definition: $b(a-i0)-b(a+i0)=2\pi i \rho_A(a)$), and this is enough to compute it.

More explicitly, by a change of variables of the form: $a\equiv P(z)$,
where $P$ is a polynomial of degree $d-1$ whose critical values are
the branching points,
we can remove all the cuts at infinity; then $b(z)$ has a single cut,
the physical cut, and it can be expressed as:
\eqn\finalbz{b(z)=c_1 z+ c_2 + \int_{z(a_1)}^{z(a_2)} {\d z'\rho_A(a(z'))\over z-z'}}
where the constants $c_1$ and $c_2$ are easily determined by
asymptotics at infinity.

The simplest non-trivial case is the cubic case, in which we find only
two sheets (the physical sheet and an extra one) if we do not cross
the physical cut, and the equation \finalbz\ reduces to the known
solution \KOGRN. Since this is the case that is of interest to us, let us
carry out explicitly the procedure outlined above.
From Eq. \spebb\ with $V(b)=-{\beta\over3}b^3+{\gamma\over2}b^2$, we
obtain that there are two sheets $b_\pm(a)$ (the physical sheet being
by definition $b_+(a)$) on which as $a\to\infty$,
\eqn\binf{b_\pm(a)=\pm (-a/\beta)^{1/2}+{\gamma\over 2\beta}
\pm {\gamma^2\over 8\beta^{3/2}} (-a)^{-1/2} + {1\over 2a} +
O(a^{-3/2})}
where we have used $\omega_B(b)=1/b+O(1/b^2)$.\foot{Had we used an
expansion $\omega_B(b)=\sum_{n=0}^\infty B_n/b^{n+1}$
where $B_n\equiv \left<{1\over N}\tr B^n\right>$,
we would have obtained an expansion at arbitrary
order of $b_\pm(a)$ in terms of the $B_n$.}
After the change of variables $z^2=a_3-a$ (where $a_3$ is the branch
point of the semi-infinite cut), we find that
$$b(z)={z\over\beta^{1/2}}+{\gamma\over 2\beta}
+{\gamma^2-4a_3\beta\over 8\beta^{3/2}}{1\over z} - {1\over 2z^2}
+O(1/z^3)$$
In particular the constants in \finalbz\ are given by 
$c_1=\beta^{-1/2}$ and $c_2={\gamma\over 2\beta}$.

Note however that in the next paragraphs, 
we shall not need to make the explicit change of variables
$a\to z$. 
It is known that this maps the saddle point equations of the $Q$-states Potts
model onto those of the $O(n)$ model \DAUL, but this correspondence has no
direct physical meaning (the phase diagrams are different,
the critical models do not
have the same central charge\rem{Technically, this comes from the
fact that the mapping $a\mapsto z$ changes by a factor of $2$ the
critical exponent $g$ related to the central charge by:
$c=1-6(\sqrt{g}-1/\sqrt{g})^2$.}, etc) and is more misleading than useful
for our purposes. Instead, we shall directly consider
$b(a)$ in its normal parametrization and
use the information on its
analytic structure discussed above.

\newsec{The $Q=1$ dilute Potts model (Ising with magnetic field)}
Let us now apply the results of the previous paragraph to the
two-matrix model. The philosophy will be the same in this case as in
the more general $Q$-states Potts model with $0<Q<4$:
we have two saddle point equations, one coming from
$A$, the other one from $B$. These equations involve functions $a(b)$
and $b(a)$ which
satisfy a functional inversion relation; therefore, by inverting one
of them, we obtain two relations for $b(a)$. Recombining
them, we deduce a polynomial equation satisfied by $b(a)$.

The saddle point equation of the eigenvalues of 
$B$ has already been analyzed in the previous
section; as we have shown, it gives the behavior of $b(a)$ at infinity
and fixes its analytic structure (Eq. \binf). We have found two
sheets, the physical sheet $b_+(a)$, and $b_-(a)$ which is connected
to it by a semi-infinite cut.

Remembering that the full partition function takes the form
$$
{\rm Z}_1(\alpha,\beta,\gamma)=\int\!\!\!\int\prod_i \d
a_i \d b_i \Delta[a_i] \Delta[b_i]
\det_{i,j}\big[\E{N a_i b_j}\big]
\E{ N\sum_i\left[-{\gamma\over 2}(a_i^2+b_i^2)
+{\alpha\over3}a_i^3+{\beta\over3}b_i^3\right]}
$$
in terms of the eigenvalues of $A$ and $B$,
we now write the analytically continued
saddle point equation for the eigenvalues of $A$:
\eqn\spea{\omega_A(a)+b_\star(a)-\gamma a+\alpha a^2=0}
$b_\star(a)$ is a third sheet of $b(a)$,
connected to the physical sheet $b_+(a)$ by the physical cut, and
which according to \spea, has no other cut than the physical cut.
This means that $b(a)$ has exactly three sheets,
and we therefore make the Ansatz that $b(a)$ is the solution of a third
degree equation in $a$. The coefficients of this polynomial are
symmetric functions of the different sheets; for example, using
the elementary identity $b_+(a)+b_-(a)=\omega_A(a)+{\gamma\over\beta}$
(coming from the analytic structure of $b(a)$)
and \spea, we find that $b_+(a)+b_-(a)+b_\star(a)=-\alpha a^2+\gamma
a+{\gamma\over\beta}$. For other symmetric functions we must use the
finite expansion \binf, which implies that
not all of the coefficients of the polynomial
can be found explicitly.
We find that there are three remaining unknowns called
$x$, $y$, $z$ (which can be reexpressed in terms of the three
first moments of $b$, cf footnote$^4$). The final equation can be cast
in the nicely symmetric form:
\eqn\polyeq{\alpha\beta\, a^2 b^2+\alpha\, a^3 
-\gamma \alpha\, a^2 b - \gamma \beta\, b^2 a + \beta\, b^3
-\gamma\, a^2 + (\gamma^2+1-\alpha\beta)\, ab -\gamma\, b^2
+x\, a +y\, b +z=0}
The constants $x$, $y$, $z$ are determined by imposing that
$a(b)$ and $b(a)$ have the appropriate analytic structure:
when solving \polyeq\ for $b$, the discriminant is a $9^{\rm th}$ degree
polynomial in $a$, and we must impose that $3$ zeroes are double
zeroes in order to have only three branch points\foot{Note that
this constraint is much stronger than the one given in \STAU\ (``one 
cut hypothesis''), which in fact would not be enough to fix the three unknowns.}.
This gives three algebraic equations for $x$, $y$, and $z$ which are therefore 
functions of $\alpha$, $\beta$, $\gamma$.

One can show that, in the absence of magnetic field ($\alpha=\beta$),
the equation \polyeq\ is equivalent to the equation
for the resolvent found in [\xref\ALF,\xref\BOUL,\xref\STAU] using
more complicated methods; however, \polyeq\ 
displays explicitly
the ${\Bbb Z}_2$ symmetry $a\leftrightarrow b$, $\alpha\leftrightarrow\beta$, whereas
it is not obvious in [\xref\ALF,\xref\BOUL,\xref\STAU].
This explicit symmetry is due to the fact that
we are not trying to write down an equation for $\omega_A(a)$ directly
(or $\omega_B(b)$), but rather for $b(a)$, which differs by a
polynomial part $-\gamma a+\alpha a^2$.

Figure \phdiag\ shows the resulting topology for
the phase diagram of the model. We shall
now explain it by briefly analyzing Eq. \polyeq\ in two particular cases.
\fig\phdiag{Schematic phase diagram of the Ising model with
magnetic field on random surfaces. \P\T\ is the zero magnetic field
low temperature phase transition line; \T\ is the critical point of the
model; \C\T\ (resp.\ \D\T) is a boundary phase transition line for
$b(a)$ (resp.\ $a(b)$).}{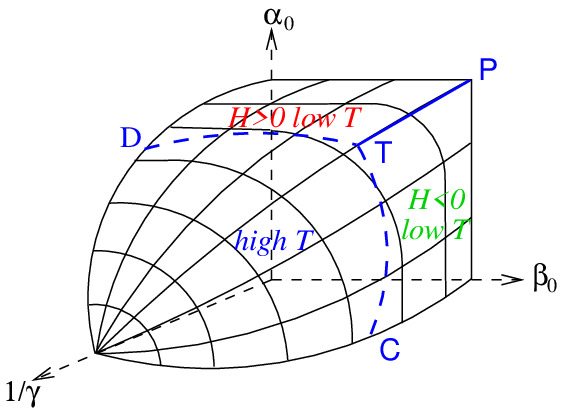}

If we first remove the dilution ($\alpha=0$, which in the Ising
language, corresponds to an infinite magnetic field), this model is simply
the one-matrix model, and we expect no physics at all; in fact,
integrating over $A$ shows that there is only one parameter in the
model, $\beta/(\gamma-1/\gamma)^{3/2}$. 
The continuum limit, which is attained for
$${\beta^2\over(\gamma-1/\gamma)^3}={1\over 12\sqrt{3}}$$
corresponds to a $c=0$ theory,
and there seems to be no critical point. However, this is {\it wrong},
because even though the bulk theory is pure gravity with a coupling
$\beta/(\gamma-1/\gamma)^{3/2}$, $b(a)$ represents a non-trivial loop
function, and the corresponding boundary operator depends explicitly
on $\gamma$. Indeed we find that the
standard resolvent $\omega_B(b)$, or equivalently
$a(b)$ (cf Eq. \speb) always has the singularity
$$a-a_{\rm reg}\sim (b-b_\star)^{3/2}$$
of the usual pure gravity loop function, but that $b(a)$ (or
$\omega_A(a)$) undergoes a phase
transition! This can be seen in the behavior of the branch points
of $b(a)$ as one varies $\gamma$ (figure \phtra). There is a critical point
(point \C\ of figure \phdiag)
$$\gamma_\c^2=1+2\sqrt{3}\qquad\beta_\c^2=2\gamma_\c^{-3}$$
where the two cuts merge, and the result is that
\eqn\singcrit{
b-b_{\rm reg}\sim\left\{\eqalign{
&(a-a_2)^{3/2}\qquad \gamma<\gamma_\c\cr
&(a-a_2)^{2/3}\qquad \gamma=\gamma_\c\cr
&(a-a_2)^{1/2}\qquad \gamma>\gamma_\c\cr
}\right.}
For $\gamma>\gamma_\c$ there is again a $(a-a_3)^{3/2}$ singularity,
but $a_3$ does not belong to the physical cut.
\fig\phtra{Cut structure of $b(a)$ for $\alpha=0$ (infinite magnetic field)
on the continuum limit surface.
The numbers represent the multiplicity of the zeroes of the discriminant.
Note that the two cuts never
intersect each other (the choice of cuts is dictated by
analytic continuation from the gaussian model).}{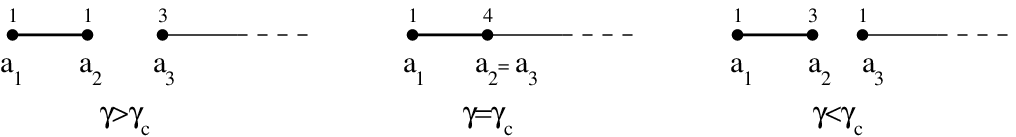}
Let us insist that 
the corresponding theory on the sphere (i.e.\ the
free energy of the matrix model) only
depends on the combination $\beta/(\gamma-1/\gamma)^{3/2}$ and is always pure
gravity; but
there is nonetheless a phase transition, at the boundary
of the surface, which is signalled by a change of analytic behavior
of $b(a)$ from $\gamma<\gamma_\c$ (high temperature) 
to $\gamma>\gamma_\c$ (low temperature) --
or equivalently,
a change in the asymptotics of $\left<\tr A^n\right>$ as $n\to\infty$.
The situation might seem trivial at the point \C\ where the integration
over $A$ is gaussian, but in fact the phase transition occurs on
a whole ``critical line'' (line \C\T\ of figure \phdiag), and
can be interpreted as follows: inside the region \P\C\T, where
the temperature is low and the magnetic field
favors spins $B$, the boundary of the random surface
is made of spins $A$, whereas the bulk of the surface is mostly
made of spins $B$, so that
the boundary tries to avoid touching the rest of the
surface by ``collapsing on itself'' (which results in a change
of its Hausdorff dimension\rem{really?}).
This is a quantum gravity phenomenon
which has no precise analogue on a flat lattice\rem{even though related
phenomena can occur at low temperature on a flat lattice with specific
boundary conditions? ref?}.
Symmetrically, $a(b)$ undergoes the same phase transition on the line
\D\T; in fact, it is clear that for any $Q\le 4$,
the point \D\ exists (its position being, up to a trivial
rescaling of $\gamma$, independent of $Q$)
and is the endpoint of a boundary phase transition
line, with the same critical properties as for $Q=1$; so that
we shall not mention this line
in the subsequent discussion of $Q=2$, $3$, $4$.

Second, let us discuss the physics on the zero magnetic field line
$\alpha=\beta$, on which the Ising critical point
(point \T\ of figure \phdiag) is.
Note that the symmetry of the equation imposes
then that $x=y$, so that there are only two
unknown constants left in \polyeq\ (as opposed to the
three found in \STAU). A very nice picture emerges out of the patterns
of the zeros of the $9^{\rm th}$ degree discriminant
(figure \ising), which in
particular illustrates the ${\Bbb Z}_2$ spontaneous symmetry breaking in the
low temperature phase.
\fig\ising{Cut structure of $b(a)$ for $\alpha=\beta$ (zero magnetic field)
on the continuum limit surface.
The numbers represent the multiplicity of the zeroes of the discriminant.
The exchange of $A$ and $B$, represented by the arrows, acts non-trivially 
on the branch points in the low temperature phase
($\gamma>\gamma_\t$), but not in the high temperature
phase ($\gamma<\gamma_\t$).}{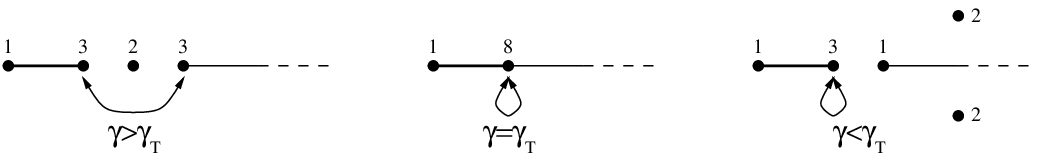}
At the critical point
$$\alpha_\t=\beta_\t=\sqrt{10}=3.16228\ldots \qquad \gamma_\t=2\sqrt{7}+1
=6.29150\ldots$$
the resolvent develops a singularity ($a_\star\equiv a_2=a_3$)
\eqn\singtri{
b-b_{\rm reg} \sim (a-a_\star)^{4/3}
}
which is characteristic of a $c=1/2$ theory coupled to gravity.
Let us note that since the resolvent must have a cubic singularity,
three is the minimum number of sheets required of $b(a)$; therefore
the cubic two-matrix model is the simplest possible realization of the
$c=1/2$ theory coupled to gravity.

Since the Ising model on random surfaces has already been studied 
in great detail, we shall not elaborate any further and go back
to the general case with arbitrary $Q$. Noting that the partition
function \Qmm\ takes the form
$${\rm Z}_Q(\alpha,\beta,\gamma)=\int\d A\, \E{N\tr\left[-{\gamma\over2}
A^2+{\alpha\over3}A^3\right]} \Xi(A)^Q$$
where $\Xi(A)$ is defined by \extf\ with the usual cubic potential,
and using \derlog, we can immediately write down
the saddle point equations
for the eigenvalues of $A$; after analytic continuation, we obtain:
\eqn\speaQ{(2-Q)\omega_A(a)+b_\star(a)+(Q-1)b(a)=\gamma a -\alpha a^2
}
where $b(a)=b_+(a)$ is evaluated on the physical sheet.
For $Q=2$ and $3$ the resolution is very similar to the $Q=1$ case,
and the resulting phase diagrams are of the same type as \phdiag,
except that the line \C\T\ becomes a real critical line (for the bulk
theory);
we shall now give the main results, skipping the technical details.

\newsec{$Q=2$ and $3$ dilute Potts models}
\rem{note to myself: $Q=2$ i.e.\ $n=0$ is the only case
where $a\mapsto z$ is really useful, since it removes the 2
semi-infinite cuts without creating any extra ``mirror image'' cuts.}
The matrix model corresponding to the
$Q=2$ dilute Potts model on random surfaces
is nothing but the ${\Bbb Z}_2$ symmetric three matrix chain \CHMAME, $A$ 
playing the role of the central matrix and $B$ of the two matrices at
the ends of the chain. If we remove the dilution, i.e.\ set
$\alpha=0$, we are back to the Ising model without magnetic field.
However, adding dilution allows us to reach the tricritical point
of the Ising model \FUHASA.

The saddle point equation \speaQ\ reads for $Q=2$:
\eqn\speaQb{b_\star(a)+b(a)=\gamma a -\alpha a^2}
which shows that $b_\star(a)$ and $b(a)$ must have the same cuts;
since we know from section 2 that $b(a)=b_+(a)$ is connected by
a semi-infinite cut to $b_-(a)$, 
$b_\star(a)=b_{\star +}(a)$ must likewise
be connected to a fourth sheet $b_{\star -}(a)$. We therefore
assume that $b(a)$ satisfies a fourth degree equation and compute
symmetric functions of its 4 sheets. Using
once again the expansion \binf, the saddle point equation \speaQb\ 
and various other relations coming
from the analytic structure of $b(a)$,
we find the following algebraic equation:
\eqn\polyeqQ{
\eqalign{
&\beta\, b^4-2\beta\big(-\alpha\, a^2+\gamma\, a\big)\, b^3
+ \big(\alpha^2\beta\, a^4 -2\alpha\beta\gamma\, a^3 + \gamma(\beta\gamma
-\alpha)\, a^2
+ (-\alpha\beta + 2+\gamma^2)\, a + \cdots\big) b^2\cr
&- \big(\gamma \alpha^2\, a^4 + \alpha(\alpha\beta-2(1+\gamma^2))\,a^3 +
\cdots\big)\, b
+ \big(\alpha^2\, a^5 - 2 \alpha\gamma\, a^4 +\cdots\big)
=0
}
}
where $\cdots$ means that there are some lower order terms whose
coefficients must be fixed by the analytic structure. Note that
as an equation for $a$, \polyeqQ\ is a fifth degree equation.

Fixing the unknown coefficients in \polyeqQ\ allows
us to easily find the critical line,
which is characterized by the collision of the physical cut with
the semi-infinite cut. Its two endpoints
are the zero dilution ($\alpha=0$) critical point:
$$
\gamma_\c^2=2(1+\sqrt{7})
\qquad
\beta_\c^2 =10\gamma_\c^{-3}
$$
where the loop functions display the same behavior that is found along
the whole critical line, namely
\eqn\singcritQ{
\eqalign{
b-b_\star&\sim (a-a_\star)^{3/4}\cr
a-a_\star&\sim (b-b_\star)^{4/3}\cr
}
}
and are characteristic of a $c=1/2$ theory;
and the tricritical point:\foot{Only the
numerical values are given, the exact values being fairly cumbersome.
This remark also applies to the cases $Q=3$ and $4$.}
$$
\alpha_\t=2.83045\ldots\qquad
\beta_\t=3.09138\ldots\qquad
\gamma_\t=6.23472\ldots
$$
where the corresponding singularities of the loop functions are
\eqn\singtriQ{
\eqalign{
b-b_{\rm reg}&\sim (a-a_\star)^{5/4}\cr
a-a_{\rm reg}&\sim (b-b_\star)^{5/4}\cr
}
}
As expected this corresponds to the central charge $c=7/10$
of tricritical Ising.

The $3$-states dilute Potts model 
is the first in which the corresponding matrix
model $Q=3$ does not have the form of a linear chain; in particular,
it is not solvable via orthogonal polynomials. We start
once more from the saddle point equation \speaQ. The discussion
of the analytic structure of $b(a)$ 
in this case becomes a little more involved,
and let us simply state the conclusion of this analysis, which is
that $b(a)$ must have 6 sheets, as shown on figure \minconjQ. The reader
can check that this analytic structure, and in particular the
discontinuities shown, are compatible with the saddle point
equations. Hence, we assume $b(a)$ is a solution
of a sixth degree equation, and look for its coefficients as
polynomials in $a$.
\fig\minconjQ{Analytic structure of $b(a)$ in the 3-states
dilute Potts model. Jumps of $b(a)$ across its cuts are 
shown.}{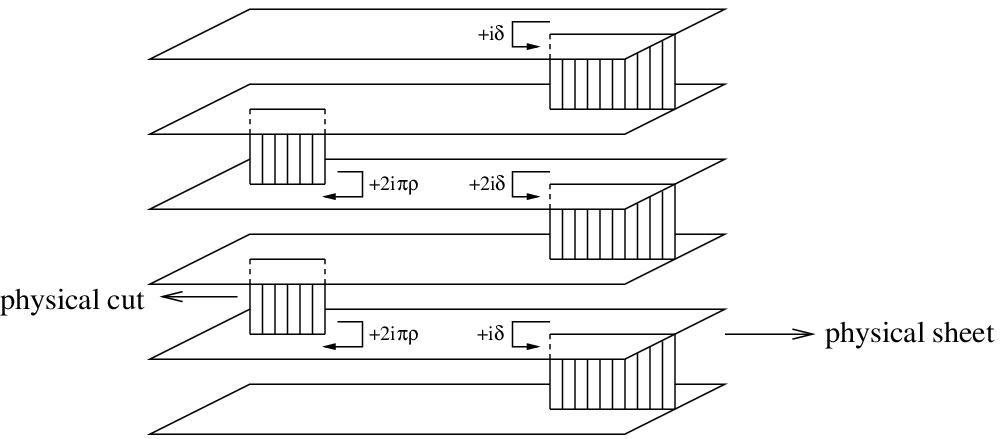}

For the general dilute $3$-states Potts model, the degree of these
polynomials becomes quite high (up to 9), and we shall only write
the algebraic equation satisfied by $a$ and $b$ in the non-dilute
($\alpha=0$) case:
\eqn\polyeqQb{
\eqalign{
&\beta\, b^6 + 6\gamma\big(1-\beta\, a\big)\, b^5
+\big(13\beta\gamma^2\,
a^2-6(-1+4\gamma^2)\,a+\gamma(2\beta+9\gamma/\beta)\big)\,b^4\cr
&+\big(-12\beta\gamma^3\, a^3+4\gamma(-6+7\gamma^2) \,a^2 - 4{\gamma\over\beta}
(-6+2\beta^2\gamma+3\gamma^2)\, a +\cdots\big)\, b^3\cr
&+\big(4\beta\gamma^4\, a^4 - 6\gamma^2 (-5+\gamma^2)\, a^3
+{1\over\beta}(9-54\gamma^2+10\beta^2\gamma^3-15\gamma^4) a^2+\cdots\big)\, b^2\cr
&+\big(-4\gamma^3(3+\gamma^2)\,a^4-{1\over\beta}(18\gamma-24\gamma^3+4\beta^2
\gamma^4-18\gamma^5)\,a^3+\cdots\big)\, b\cr
&+\big(4\gamma^4\, a^5 +{\gamma^2\over\beta}(17-18\gamma^2)\, a^4
+\cdots\big)=0\cr}
}
Note that it was not known before that the standard resolvent of the
Potts model $\omega_B(b)$ (or
$a(b)$, cf Eq. \speb) satisfies a polynomial equation.
This a fifth degree equation in $a$, so that $a(b)$ has five sheets.
The critical point is easily determined:
$$\gamma_\c^2=3+\sqrt{47}\qquad\beta_\c^2={105\over 4}\gamma_\c^{-3}$$
which is compatible with what was found in \DAUL. The singularity of the
resolvents is 
\eqn\singcritQb{\eqalign{
b-b_\star&\sim(a-a_\star)^{5/6}\cr
a-a_\star&\sim(b-b_\star)^{6/5}\cr
}}
which corresponds to a $c=4/5$ theory.

If we introduce the dilution, the equations become rather complicated, though
it is still possible to work with exact analytic expressions. We find a
critical line, 
with singularities of the type \singcritQb, which ends with
a tricritical point:
$$\alpha_\t=2.44405\ldots\qquad \beta_\t=2.9536\ldots\qquad\gamma_\t=6.09718\ldots$$
at which the singularity becomes
\eqn\singtriQb{
\eqalign{
b-b_{\rm reg}&\sim(a-a_\star)^{7/6}\cr
a-a_{\rm reg}&\sim(b-b_\star)^{7/6}\cr
}
}
that is a $c=6/7$ theory.

\newsec{$Q=4$ dilute Potts model}
For the $Q=4$ case, it is easy to show that $b(a)$ has an infinite
number of sheets; of course, the method used above for $Q<4$ then fails.
In order to solve the model, the easiest procedure is to follow
\PZJK\ in which a similar ``double saddle point equation'' system was
solved.
One introduces an auxiliary function 
$$D(a)\equiv 2b(a)-\omega_A(a)-{\gamma\over\beta}$$
which, for $Q=4$ (and only $Q=4$),
satisfies a two-cut Riemann--Hilbert problem:
\eqn\twoRH{\matrix{\displaystyle
D(a+i0)+D(a-i0)=-\alpha a^2+\gamma
a-{2\gamma\over\beta}\hfill& a\in[a_1,a_2]\cr
\displaystyle
D(a+i0)+D(a-i0)=0\hfill& a\in[a_3,+\infty]\cr
}}
where $[a_1,a_2]$ is the physical cut and $[a_3,+\infty]$ is the
semi-infinite cut.
$D(a)$ can therefore be expressed via
an elliptic parametrization in terms of $\Theta$-functions.
The critical line is expected to be found (just as in \PZJK) when
the two cuts meet ($a_2=a_3$), that is in the trigonometric limit
of the elliptic functions. The solution of
\twoRH\ becomes then much simpler, and is of the form:
$$D(a)=-{\alpha\over\pi}(a-a_2)\left((a-a_0)
\arctan\sqrt{a_2-a_1\over a_1-a}+
\sqrt{(a_2-a_1)(a_1-a)}\right)$$
where $a_1$, $a_2$, $a_0$ and the relation defining the critical line are
given by the asymptotics of $D(a)$ as $a\to\infty$ and the
condition $-\alpha(a-a_2)(a-a_0)
=-\alpha a_2^2+\gamma a_2-{2\gamma\over\beta}$.

It is easy to see that positivity of the density of eigenvalues
implies that $a_0\ge a_2$. In particular for $\alpha=0$
($a_0=+\infty$), one finds
$$\gamma_\c^2=4(1+\sqrt{1+\pi^2/3})
\qquad\beta_\c^2={16\pi^2\over 3}\gamma_\c^{-3}$$
More generally, along the critical line one has the
strict inequality $a_0>a_2$
and the resolvent has the singularity
\eqn\singcritQc{
b-b_{\rm reg}\sim (a-a_\star)\log^2(a-a_\star)
}
or after inversion
\eqn\singcritQd{
a-a_{\rm reg}\sim {b-b_\star\over\log^2(b-b_\star)}
}
On the other hand, at the tricritical point
$$\alpha_\t=1.6523\ldots\qquad\beta_\t=2.42087\ldots
\qquad\gamma_\t=5.4602\ldots$$
one has $a_0=a_2$, and
this reflects in a change of the leading singularity:
\eqn\singtriQc{
b-b_{\rm reg}\sim (a-a_\star)\log(a-a_\star)
}
or
\eqn\singtriQd{
a-a_{\rm reg}\sim {b-b_\star\over\log(b-b_\star)}
}
The behavior \singcritQc\ and \singtriQc\ of the resolvent
$b(a)$ is the usual one and is also found in other $c=1$ models like the
$O(2)$ model [\xref\KOS,\xref\KOSTA]; however,
the loop functions \singcritQd\ (in particular the standard resolvent
$a(b)$ of the non-dilute $4$-states Potts model) and
\singtriQd\ are more unusual,
and should be compared with the similar result found in \PZJK.

\newsec{Conclusion}
We have discussed a new method for solving multi-matrix models
in the large $N$ limit,
which relies on saddle point analysis
and functional inversion relations satisfied by the
unknown functions appearing in the saddle point equations. This method
gives a new way of deriving equations for the resolvents of the model,
and is a (much simpler) alternative to the method of loop equations
for the determination of the disk amplitudes.
It is not limited, like the orthogonal polynomials, to
linear chains of matrices, as we have shown by solving explicitly
the $(Q+1)$-matrix model representing the dilute Potts model
on random surfaces.

Furthermore,
we have seen that the functional inversion can be interpreted here
as a duality of the theory. In the Ising model this is simply
the ${\Bbb Z}_2$ symmetry of the model; but more generally,
in the dilute Potts model, the functional inversion relates 
different boundary conditions for surfaces
with the topology of a disk. On the critical line, it exchanges
two different loop functions with different critical exponents 
($g\to 1/g$, while the bulk theory is unchanged, cf \cc) --
similarly to the duality that exchanges Dirichlet and Neumann 
boundary conditions in open string theory. At the $Q<4$
tricritical points, the universal part of the loop function
turned out to be self-dual, whereas logarithmic corrections
spoil the self-duality at $Q=4$.

One can hope that the ideas presented here
will be applicable to various other problems
of statistical mechanics on random surfaces, combinatorics, and
asymptotics of orthogonal polynomials.

\centerline{\bf Notes added}
After this work was completed, I became aware of the
fact the critical point with singularities given by Eq.~\singcrit\ is also
the critical point of a percolation model, as
noticed first in \ref\KAZp{V.A.~Kazakov,
{\it Mod. Phys. Lett.} A4 (1989), 1691.} (I thank D.~Jonhnston and
M.~Stathakopoulos for pointing this out to me). Also, it was
claimed in a recent preprint \ref\GAB{G.~Bonnet,
preprint {\tt hep-th/9904058}.} that using loop equations
one could reproduce the polynomial equation \polyeqQb\ of
the 3-states Potts model.

\centerline{\bf Acknowledgements}
I thank I.~Kostov for e-mail discussions.
This work was supported in part by the DOE grant DE-FG02-96ER40559.

\listrefs
\bye